\documentclass[fleqn,usenatbib]{mnras}

\usepackage{amssymb}

\usepackage{newtxtext,newtxmath}
\usepackage{xcolor}

\usepackage[T1]{fontenc}
\usepackage{gensymb}

\DeclareRobustCommand{\UD}[3]{#2}
\let\UDthebibliography\thebibliography
\def\thebibliography{\DeclareRobustCommand{\UD}[3]{##3}\UDthebibliography}

\usepackage{graphicx}	
\usepackage{amsmath}	
\usepackage{amssymb}	
\usepackage{pdflscape}	
\usepackage{hyperref}

\newcommand{\sigorie}{$\sigma$\,Ori\,E}
\newcommand{\bz}{$\langle B_{\rm z}\rangle$}

\newcommand\bzmin{$\langle B_{\rm z}\rangle_{\rm min}$}
\newcommand\bzmax{$\langle B_{\rm z}\rangle_{\rm max}$}

\title[Tr16-26]{Trumpler 16-26: A New Centrifugal Magnetosphere Star Discovered via SDSS/APOGEE $H$-band Spectroscopy\footnote{Based on observations made with the Southern African Large Telescope (SALT)}}

\author[S. Drew Chojnowski et al.]{
S. Drew Chojnowski$^{1}$,\thanks{E-mail: stephen.chojnowski@montana.edu}
Swetlana Hubrig$^{2}$,
Jonathan Labadie-Bartz$^{3}$,
Thomas Rivinius$^{4}$,
Markus Sch\"{o}ller$^{5}$,
\newauthor Ewa Niemczura$^{6}$,
David L. Nidever$^{1}$,
Amelia M. Stutz$^{7}$,
and C.A. Hummel$^{8}$
\\
$^{1}$Department of Physics, Montana State University, P.O. Box 173840, Bozeman, MT 59717-3840\\
$^{2}$Leibniz-Institut f\"{u}r Astrophysik Potsdam (AIP), An der Sternwarte 16, 14482 Potsdam, Germany\\
$^{3}$Instituto de Astronomia, Geof\'{i}sica e Ci\^{e}ncias Atmosf\'{e}ricas, Universidade de S\~{a}o Paulo, Rua do Mat\~{a}o 1226, Cidade Universit\'{a}ria, 05508-900 S\~{a}o Paulo, SP, Brazil\\
$^{4}$European Organisation for Astronomical Research in the Southern Hemisphere (ESO), Casilla 19001, Santiago 19, Chile\\
$^{5}$European Southern Observatory, Karl-Schwarzschild-Str.~2, 85748 Garching, Germany\\
$^{6}$Instytut Astronomiczny, Uniwersytet Wroc\l{}awski, Kopernika 11, 51-622 Wroc\l{}aw, Poland\\
$^{7}$Departmento de Astronom\'{i}a, Universidad de Concepci\'{o}n, Casilla 160-C, 4030000 Concepci\'{o}n, Chile\\
$^{8}$European Southern Observatory, Karl-Schwarzschild-Str.~2, 85748 Garching, Germany
}

\date{Accepted XXX. Received YYY; in original form ZZZ}

\pubyear{2022}

\begin{document}
\label{firstpage}
\pagerange{\pageref{firstpage}--\pageref{lastpage}}
\maketitle

\begin{abstract}
We report the discovery of a new example of the rare class of highly magnetized, rapidly rotating, helium enhanced, early B stars that produce anomalously wide hydrogen emission due to a centrifugal magnetosphere (CM). The star is Trumpler 16-26, a B1.5\,V member of the Trumpler 16 open cluster. A CM was initially suspected based on hydrogen Brackett series emission observed in SDSS/APOGEE $H$-band spectra. Similar to the other stars of this type, the emission was highly variable and at all times remarkable due to the extreme velocity separations of the double peaks (up to 1300\,km\,s$^{-1}$.) Another clue lay in the TESS lightcurve, which shows two irregular eclipses per cycle when phased with the likely 0.9718115 day rotation period, similar to the behavior of the well known CM host star {\sigorie}. To confirm a strong magnetic field and rotation-phase-locked variability, we initiated a follow-up campaign consisting of optical spectropolarimetry and spectroscopy. The associated data revealed a longitudinal magnetic field varying between $-3.1$ and $+1.6$\,kG with the period found from photometry. The optical spectra confirmed rapid rotation ($v$\,sin\,$i=195$\,km\,s$^{-1}$), surface helium enhancement, and wide, variable hydrogen emission. Tr16-26 is thus confirmed as the 20$^{\rm th}$ known, the fourth most rapidly rotating, and the faintest CM host star yet discovered. With a projected dipole magnetic field strength of $B_{\rm d}>11$\,kG, Tr16-26 is also among the most magnetic CM stars.
\end{abstract}




\section{Introduction} \label{intro}

Despite being expected to lack surface convection zones in which the magnetic fields of low-mass stars are generated, observations over the past few decades have confirmed that a non-negligible fraction of early-B stars are in fact hosts to strong, kilo-Gauss (kG) magnetic fields. Perhaps the best known example is {\sigorie}, a rapidly rotating ($v \sin\ i\approx150$\,km\,s$^{-1}$) B2\,V star whose He-strong nature was noted in the literature as early as \citet{1958ApJ...127..237G}. The strong magnetic field of {\sigorie} was first reported by \citet{1978ApJ...224L...5L}, who found the longitudinal field strength to vary between roughly $-2.2$ and $+2.5$\,kG over the course of the 1.19\,day rotation period. When treated under the framework of the oblique dipole rotator model \citep{1967ApJ...150..547P}, the implication is a dipole magnetic field strength of $B_{\rm d}\approx10$\,kG. More recent studies indicate that {\sigorie}'s magnetic field is more complicated than a simple dipole, but that it can be approximated by a dipole component with $B_{\rm d}\sim7.5$\,kG and a quadrupole component with $B_{\rm q}\sim4$\,kG \citep{2015MNRAS.451.2015O}. Regardless of the details, {\sigorie} is one of the most magnetic main sequence stars known. Detection of strong magnetic fields for additional He-strong and He-weak B stars \citep[e.g.][]{1987ApJ...323..325B} has shown that, similar to the classical Ap/Bp stars, the chemical peculiarities of stars like {\sigorie} are a result of magnetism.

An additional peculiarity of {\sigorie} and an exceedingly limited number of other magnetic early-B stars is the presence of extremely wide, double peaked emission in the hydrogen lines, with the morphology of the double peaked line profiles varying exactly according to rotational phase, $\phi_{\rm rot}$ \citep{1993AAS...182.4506B}. The bulk of emission in each line occurs well beyond $v \sin\ i/W$, which is not possible in Keplerian disks like those of classical Be stars \citep[where $W$ is the rotation parameter defined as the ratio of the equatorial over orbital velocity, $W=v_{\rm eq}/v_{\rm orb}$;][]{2013A&ARv..21...69R}. Rather, the circumstellar emission of {\sigorie} types can be understood in the context of the rigidly rotating magnetosphere (RRM) model of \citet{2005MNRAS.357..251T}, whereby stellar wind material accumulates at the intersections of the magnetic and rotational axes and is forced by the magnetic field to rotate at the same angular velocity as the surface of the central star. 

Thanks to recent spectropolarimetric surveys of OB stars, the known sample of magnetic massive stars (i.e. stars with effective temperatures, $T_{\rm eff}>16$\,kK) has grown significantly over the past several decades to the point where almost 100 examples are known by now. These stars account for a non-negligible $\sim$7--10\% of the OB stars that have been observed with modern spectropolarimeters \citep{2012AIPC.1429...67G, 2017MNRAS.465.2432G}. The sample size was just 64 in 2013, but this was already sufficient for the development of a massive magnetic star classification scheme \citep{2013MNRAS.429..398P} based on estimates of the Alfv\'{e}n radius ($R_{\rm A}$), below which stellar winds can be trapped by the magnetic field, and the Kepler co-rotation radius ($R_{\rm K}$), above which the centrifugal force exceeds the gravitational force and prevents the wind material from falling back onto the star \citep{2008MNRAS.385...97U}. 

In the case of $R_{\rm K}<R_{\rm A}$, also known as a centrifugal magnetosphere (CM), the magnetically confined wind material is forced to co-rotate with the star in the region between $R_{\rm K}$ and $R_{\rm A}$. Given sufficient density of the gas, the characteristic wide hydrogen emission will arise and vary in strength and morphology as a function of $\phi_{\rm rot}$. In the region below $R_{\rm K}$, magnetically confined winds receive insufficient centrifugal support and thus will fall back onto the stellar surface over a dynamical timescale. This is referred to as a dynamical magnetosphere (DM). Whereas all magnetic stars have a DM, only those with sufficiently fast rotation and sufficiently strong magnetic fields have a CM. A total of 18 CM stars were known in 2020 and the sample was scrutinized in a series of papers by \citet{2018MNRAS.475.5144S, 2019MNRAS.485.1508S, 2019MNRAS.490..274S, 2020MNRAS.499.5379S}. 

This paper focuses on the discovery of a new example of a CM star with the aforementioned remarkable properties: Trumpler\,16-26 (hereafter referred to as Tr16-26), a B1.5\,V star that resides on the sky just 133{\arcsec} from the famous luminous blue variable $\eta$\,Car. Given the faintness of Tr16-26 ($m_{V}\approx11.55$), it is not surprising that literature mentions of the star are limited. However, numerous membership studies \citep[e.g.,][]{1993AJ....105.1822C, 2014A&A...564A..79D} have shown that Tr16-26 is almost certainly a member of the Trumpler\,16 open cluster. Similar to many of the known CM stars (e.g. {\sigorie}, HR\,5907, HD\,164492C, HD\,37017, etc.) and as expected for massive stars with magnetically confined winds \citep{1997A&A...323..121B, 2014ApJS..215...10N}, Tr16-26 is an x-ray source, having been detected by the Chandra satellite \citep{2011ApJS..194....2B}. The Chandra detection let to inclusion of Tr16-26 among a sample of Carina Nebula OB stars whose stellar parameters were investigated by \citet[][H18 from here on]{2018AJ....155..190H} based on blue optical spectra that did not cover H$\alpha$ nor H$\beta$, where wide emission would have been evident. The stellar parameter estimates were used to infer a distance of $7390\pm3870$\,pc, such that Tr16-26 would be a background giant/subgiant rather than a member of the Trumpler\,16 cluster.

More recently, parallaxes from the early third data release \citep[eDR3;][]{2020yCat.1350....0G} of the European Space Agency (ESA) Gaia mission \citep{2016A&A...595A...1G} were used to estimate a distance to the Trumpler\,16 cluster of 2320\,pc \citep{2021ApJ...914...18S}. For Tr16-26 itself, the Gaia eDR3 proper motions ($\mu_{\rm tot}=7.3$\,mas\,yr$^{-1}$) and distance \citep[2261\,pc][]{2021AJ....161..147B} are both consistent with Trumpler 16 cluster membership, such that the inferred stellar parameters reported by H18 cannot be correct. Rather, we use a combination of $H$-band and optical spectroscopy, optical photometry, and optical spectropolarimetry to argue that Tr16-26 is a main sequence CM host analogous to {\sigorie} and the faintest member of this group yet discovered. 

An overview of the available data for Tr16-26 is provided in Section~\ref{data}. In Section~\ref{prot}, we use the lightcurve from the Transiting Exoplanet Survey Satellite \citep[TESS;][]{2015JATIS...1a4003R} to refine the $\sim1$\,day rotation period of Tr16-26 that had already been estimated via photometry from All-Sky Automated Survey for Supernovae \citep[ASAS-SN;][]{2014AAS...22323603S,2019MNRAS.485..961J}. In Section~\ref{bfield}, we analyze the longitudinal magnetic field measurements and establish a lower limit on the field strength. In Section~\ref{spectype}, we revise the stellar parameters of Tr16-26 to make them consistent with both membership in the Trumpler\,16 cluster and with the stellar parameter range of the known CM stars. Finally, we investigate the spectroscopic variability and hydrogen emission in Section~\ref{specvar}.

\begin{table*}
\caption{Summary of observations and magnetic field measurements.\label{tab:obs}}
\begin{tabular}{lccccrcrrr}
\hline
Star  &  Facility  &  UT Date  &  JD$-$2.4E6  &  $t_{\rm exp}$ (s)  &  S/N  &  $\phi_{\rm rot}$  &  $\langle B_{\rm z}\rangle_{\rm all}$ (G)  &  $\langle B_{\rm z}\rangle_{\rm hyd}$ (G) & $\langle N_{\rm z}\rangle_{\rm all}$ (G) \\ 
\hline
Tr16-26 & LCO/APOGEE & 2019-06-19T23:36:60 & 58654.4840 &    2002 &      71 & 0.292 &           ... &           ... &           ... \\ 
 & LCO/APOGEE & 2020-02-04T08:38:04 & 58883.8598 &    2002 &      71 & 0.321 &           ... &           ... &           ... \\ 
 & LCO/APOGEE & 2020-03-06T04:29:18 & 58914.6870 &    5005 &     120 & 0.043 &           ... &           ... &           ... \\ 
 & LCO/APOGEE & 2020-12-04T08:38:56 & 59187.8604 &    1001 &      63 & 0.140 &           ... &           ... &           ... \\ 
 & LCO/APOGEE & 2020-12-19T08:02:46 & 59202.8353 &    2002 &      92 & 0.549 &           ... &           ... &           ... \\ 
 & LCO/APOGEE & 2021-01-09T08:24:37 & 59223.8504 &    2002 &      75 & 0.174 &           ... &           ... &           ... \\ 
 & SALT/HRS   & 2021-03-06T08:41:14 & 59279.5132 &    2200 &     121 & 0.451 &           ... &           ... &           ... \\ 
 & SALT/HRS   & 2021-03-07T08:43:40 & 59280.3152 &    2200 &     112 & 0.276 &           ... &           ... &           ... \\ 
 & SALT/HRS   & 2021-03-08T08:52:26 & 59281.3457 &    2200 &     114 & 0.337 &           ... &           ... &           ... \\ 
 & SALT/HRS   & 2021-03-09T09:49:35 & 59282.5119 &    2200 &     115 & 0.537 &           ... &           ... &           ... \\ 
 & VLT/UVES   & 2020-12-11T05:08:09 & 59194.7306 &    2600 &      96 & 0.209 &           ... &           ... &           ... \\ 
 & VLT/FORS2  & 2020-12-23T05:10:54 & 59206.7159 & $8\times509$ &     984 & 0.542 & $-2351\pm297$ & $-2409\pm411$ &    $98\pm312$ \\ 
 & VLT/FORS2  & 2020-12-25T04:42:58 & 59208.6965 & $8\times509$ &    1358 & 0.580 & $-2436\pm272$ & $-2535\pm390$ &    $52\pm295$ \\ 
 & VLT/FORS2  & 2020-12-26T04:39:48 & 59209.6943 & $8\times509$ &    1473 & 0.607 & $-2558\pm256$ & $-2946\pm374$ &  $-249\pm269$ \\ 
 & VLT/FORS2  & 2020-12-27T05:33:04 & 59210.7313 & $8\times509$ &    1905 & 0.674 & $-1647\pm166$ & $-1342\pm268$ &   $133\pm190$ \\ 
 & VLT/FORS2  & 2020-12-28T04:29:34 & 59211.6872 & $8\times509$ &    1464 & 0.658 & $-1849\pm241$ & $-1493\pm331$ &    $77\pm259$ \\ 
 & VLT/FORS2  & 2020-12-28T05:56:50 & 59211.7478 & $8\times509$ &    2115 & 0.720 &   $459\pm146$ &   $423\pm271$ &   $167\pm168$ \\ 
 & VLT/FORS2  & 2021-02-04T03:00:00 & 59249.6250 & $8\times509$ &    1485 & 0.696 &  $-204\pm185$ &  $-115\pm301$ &  $-273\pm201$ \\ 
 & VLT/FORS2  & 2021-02-23T04:56:47 & 59268.7061 & $8\times509$ &    2033 & 0.330 & $-1182\pm154$ & $-1587\pm221$ &   $223\pm175$ \\ 
 & VLT/FORS2  & 2021-02-24T01:36:03 & 59269.5667 & $8\times509$ &    1412 & 0.216 &  $1428\pm208$ &  $1362\pm325$ &   $-69\pm204$ \\ 
 & VLT/FORS2  & 2021-02-25T06:04:11 & 59270.7529 & $8\times509$ &    1988 & 0.437 & $-2581\pm191$ & $-2945\pm284$ &  $-236\pm207$ \\ 
 & VLT/FORS2  & 2021-03-01T03:58:45 & 59274.6658 & $8\times509$ &    2054 & 0.463 & $-2703\pm173$ & $-2910\pm277$ &   $-50\pm191$ \\ 
 & VLT/FORS2  & 2021-03-05T03:25:03 & 59278.6424 & $8\times509$ &    2212 & 0.555 & $-3142\pm207$ & $-3053\pm327$ &  $-178\pm200$ \\ 
 & VLT/FORS2  & 2021-03-05T07:26:15 & 59278.8099 & $8\times509$ &    1993 & 0.727 &   $421\pm250$ &   $460\pm319$ &   $-61\pm273$ \\ 
 & VLT/FORS2  & 2021-03-06T08:02:41 & 59279.8352 & $8\times509$ &    2517 & 0.782 &  $1598\pm235$ &  $1846\pm370$ &  $-120\pm244$ \\ 
 & VLT/FORS2  & 2021-03-08T03:46:05 & 59281.6570 & $8\times509$ &    1420 & 0.657 & $-1453\pm233$ & $-1430\pm329$ &   $144\pm229$ \\ 
 & VLT/FORS2  & 2021-03-09T01:13:00 & 59282.5507 & $8\times509$ &    1561 & 0.577 & $-2831\pm195$ & $-2982\pm291$ &   $138\pm191$ \\ 
\hline
HD\,23478   &  APO/ARCES   &  2014-01-14T00:46:57  &  56671.5326  &   450  &  125  &  0.805  & ... & ... & ... \\ 
            &  APO/APOGEE  &  2017-11-04T08:56:04  &  58061.8723  &  2503  &  572  &  0.190  & ... & ... & ... \\ 
            &  APO/APOGEE  &  2017-11-28T07:15:26  &  58085.8024  &  2002  &  699  &  0.985  & ... & ... & ... \\ 
\hline
HD\,345439  &  APO/ARCES   &  2012-09-02T07:40:38  &  56172.8199  &   900  &   75  &  0.333  & ... & ... & ... \\ 
            &  APO/APOGEE  &  2011-09-03T03:57:35  &  55807.6650  &  1001  &   70  &  0.450  & ... & ... & ... \\ 
            &  APO/APOGEE  &  2018-04-06T11:29:26  &  58214.9788  &  2002  &   85  &  0.201  & ... & ... & ... \\ 
\hline
{\sigorie}  &  APO/ARCES   &  2017-02-11T03:12:55  &  57795.6340  &   300  &  220  &  0.160  & ... & ... & ... \\ 
            &  APO/APOGEE  &  2016-02-18T02:10:11  &  57436.5904  &  2503  &  648  &  0.651  & ... & ... & ... \\ 
            &  APO/APOGEE  &  2017-02-10T03:56:17  &  57794.6641  &  2002  &  499  &  0.346  & ... & ... & ... \\ 
\hline
\multicolumn{10}{l}{Note: S/N values were measured around 5500\,{\AA} from echelle data, around 5500\,{\AA} from VLT/FORS2 data, and around 16000\,{\AA} from APOGEE data.} \\ 
\end{tabular}
\end{table*}

\section{Observations} \label{data}

\subsection{APOGEE $H$-band Spectroscopy}
The Apache Point Observatory Galactic Evolution Experiment \citep[APOGEE;][]{2017AJ....154...94M} is one of the sub-surveys of the Sloan Digital Sky Survey \cite[SDSS; e.g.][]{2020ApJS..249....3A} that has been operating on the Sloan 2.5-m telescope \citep{2006AJ....131.2332G} at Apache Point Observatory (APO, APOGEE-N) since 2011, and on the Ir\'{e}n\'{e}e du Pont 2.5-m telescope at Las Campanas Observatory (LCO, APOGEE-S) since 2017. The APOGEE instruments are duplicate 300-fiber, $R\approx22\,500$ spectrographs \citep{2019PASP..131e5001W} that record most of the $H$-band (15145--16960 {\AA}; vacuum wavelengths used throughout this paper when referring to the $H$-band) onto three detectors, with gaps between 15800--15860 {\AA} and 16430--16480 {\AA} due to non-overlapping wavelength coverage of the detectors. For a detailed description of the APOGEE data reduction pipeline, see \citet{2015AJ....150..173N}. All of the APOGEE spectra used in this paper were made publicly available in SDSS DR17 \citep{2022MNRAS.509.4024D}.

Tr16-26 was observed by the APOGEE-S instrument six times between 2019 June 19 and 2021 January 9. Total exposure times ranged from 17--83 minutes, and the associated signal-to-noise ratios ($S/N$) range from 63 to 120, with most being around 70. It is worth noting that due to the large  diameter (2{\arcsec}) on-sky field of view of the APOGEE fibers as well as to the location of Tr16-26 in the Carina Nebula, the spectra captured weak nebular emission in the hydrogen Brackett series lines. Fortunately, the nebular emission is exceedingly narrow with respect to the circumstellar emission, such that there is no confusing the two. 

For comparison to known CM stars, APOGEE spectra of {\sigorie}, HD\,23478, and HD\,345439 are displayed and discussed in this paper. These stars have been observed by the APOGEE-N instrument 10, 26, and 5 times, respectively, and the spectra of HD\,23478, and HD\,345439 have already been discussed by \citet{2014ApJ...784L..30E} and \citet{2015ApJ...811L..26W}. We only make use of the maximum and minimum emission spectra.

All of the spectroscopic observations used in this paper are summarized in Table~\ref{tab:obs}. The first eight columns give the star names, the facility/instruments used, the UT and Julian dates (JD) at the middle of the exposures (or in the case of FORS\,2, the middle of the eight sub-exposures), the exposure times ($t_{\rm exp}$), the $S/N$ per spectral resolution element, and $\phi_{\rm rot}$ at the times of observation. The final three columns provide the magnetic field measurements (see Section~\ref{fors2}). The quoted $\phi_{\rm rot}$ were calculated using periods and emphemerides established in this paper for Tr16-26, and taken from the literature for {\sigorie} \citep{2010ApJ...714L.318T}, HD\,23478 \citep{2015MNRAS.451.1928S}, and HD\,345439 \citep{2015ApJ...811L..26W,2017MNRAS.467L..81H}. In all cases, the emphemerides have been adjusted such that $\phi_{\rm rot}=0$ corresponds to an epoch of minimum {\bz} (as will be the convention in this paper).

\subsection{ARC/ARCES Optical Spectroscopy}
In addition to the APOGEE comparison spectra, this paper also makes use of optical spectra of {\sigorie}, HD\,23478, and HD\,345439 from the Astrophysical Research Consortium (ARC) 3.5\,m telescope and its \'{e}chelle spectrograph \citep[ARCES;][]{2003SPIE.4841.1145W}. ARCES records most of the optical wavelength range (3500--10400\,{\AA}) at a resolution of $R\approx31\,500$. The data were reduced using tasks included in the Image Reduction and Analysis Facility (IRAF\footnote{IRAF is distributed by the National Optical Astronomy Observatories, which are operated by the Association of Universities for Research in Astronomy, Inc.}).

\subsection{SALT/HRS Optical Spectroscopy}
Tr16-26 was observed four times by the High Resolution Spectrograph \citep[HRS;][]{2010SPIE.7735E..4FB} on the 9.2-m South African Large Telescope \citep[SALT;][]{2006SPIE.6267E..0ZB} at the South African Astronomical Observatory between 2021 March 6 and 2021 March 9. HRS is an \'{e}chelle spectrograph that records the optical spectrum (3700-8900\,{\AA}) onto two CCDs via a dichroic beam splitter. The observations were carried out in medium resolution mode ($R_{\rm blue}\approx43\,400$, $R_{\rm red}\approx39\,600$). The SALT/HRS spectra were calibrated and reduced using the SALT HRS MIDAS pipeline, based on the ESO MIDAS pipeline \citep{2016MNRAS.459.3068K, 2017ASPC..510..480K}. The standard steps of over-scan correction, bias subtraction, scattered light extraction, division by a normalized flat-field, subtraction, definition and extractions of \'{e}chelle orders and wavelength calibration by a ThAr lamp were applied.

Due to the relatively large field of view of the science fiber (2$\farcs$2), the HRS spectra of Tr16-26 captured significant contributions from the Carina Nebula. Two sets of nebular lines (separated by 44\,km\,s$^{-1}$) are present in each spectrum, with the species including the hydrogen Balmer series, [O~{\sc i}], [O~{\sc iii}], [N~{\sc ii}], [S~{\sc ii}], He~{\sc i}, etc. 

\subsection{VLT/UVES Optical Spectroscopy}
Tr16-26 was observed by the UV-visual \'{e}chelle spectrograph \citep[UVES;][]{2000SPIE.4008..534D} on the 8.2-m Kueyen unit of the Very Large Telescope (VLT) at the Cerro Paranal Observatory on 2020 December 11. UVES uses a beam splitter to record the full optical wavelength range (3000--11000\,{\AA}) onto two CCDs, achieving resolutions of $R\approx80\,000$ in the blue when using a 0$\farcs$4 slit and $R\approx110\,000$ in the red when using a 0$\farcs$3 slit. The UVES data were reduced using version 6.1.3 of the European Southern Observatory (ESO) Reflex software. 

\begin{figure*}
\includegraphics[width=\textwidth]{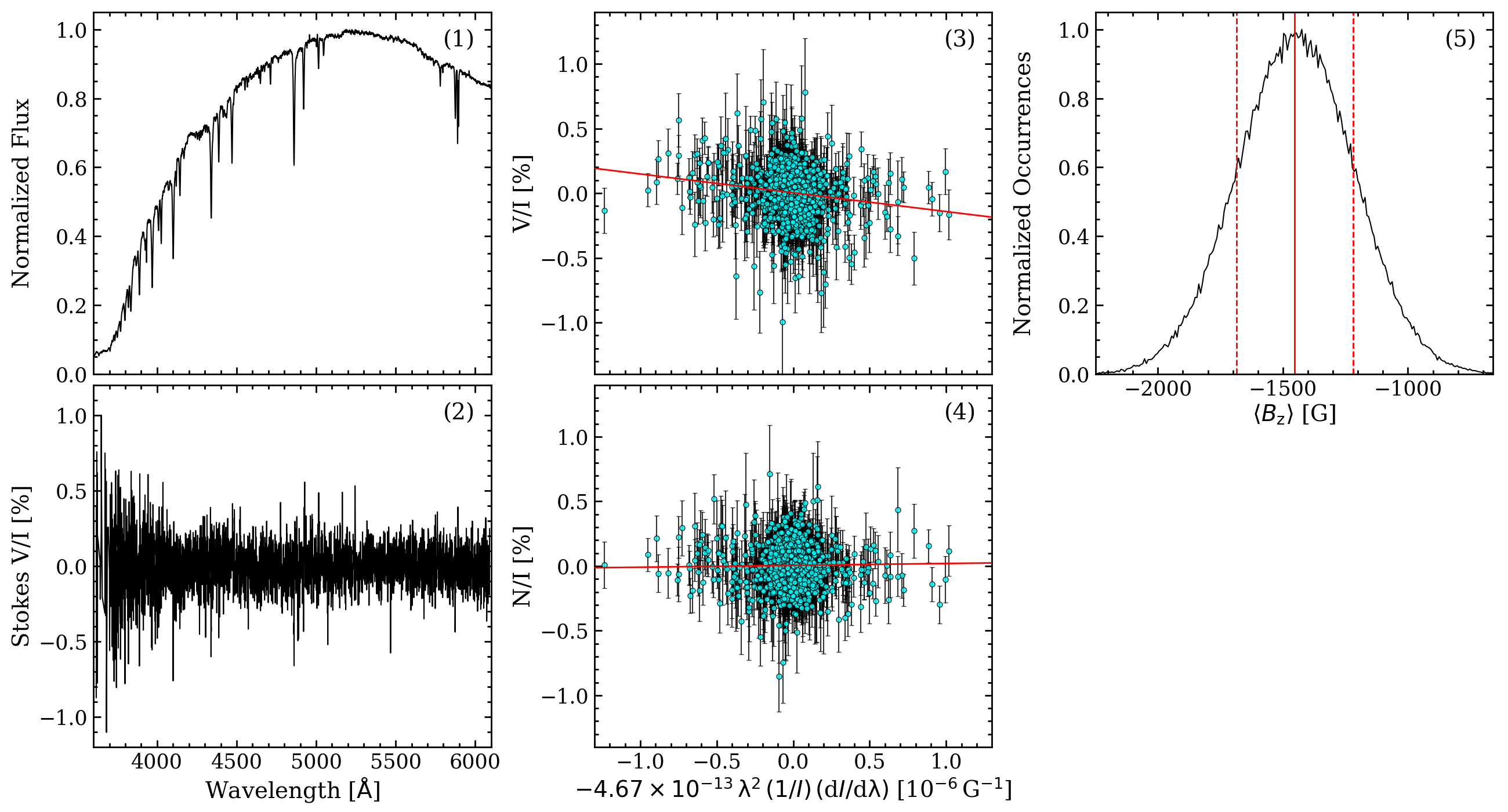}
\caption{Overview of the results of the analysis of the FORS\,2 data of Tr16-26, collected during the night of 2021 March 8, considering the full spectrum. {\sl (1)} Observed Stokes~$I$ spectrum arbitrarily normalized to the highest value. {\sl (2)} Stokes~$V$ profile (in \%). {\sl (3)} Linear fit to Stokes~$V$. {\sl (4)} Linear fit to the $N$ spectrum. From the linear fit, we determine $\left<N_{\rm z}\right> = 144\pm229$\,G. {\sl (5)} Distribution of the longitudinal magnetic field values $P(\left<B_{\rm z}\right>)$, which were obtained via bootstrapping. From the distribution $P(\left<B_{\rm z}\right>)$, we obtain the most likely value for the longitudinal magnetic field $\left<B_{\rm z}\right> = -1453\pm233$\,G. }
\label{fig:MFs}
\end{figure*}

\subsection{VLT/FORS\,2 Optical Spectropolarimetry} \label{fors2}
Tr16-26 was observed 16 times in service mode by the FOcal Reducer/low dispersion Spectrograph 2 (FORS\,2) mounted on the 8.2-m Antu unit of the VLT between 2020 December 23 and 2021 March 9. The FORS\,2 instrument is equipped with polarization analyzing optics, comprising super-achromatic half-wave and quarter-wave phase retarder plates, and a Wollaston prism with a beam divergence of 22$\arcsec$ in standard resolution mode. Polarimetric spectra were obtained with the GRISM~600B and a slit width of 0$\farcs$5 to achieve a spectral resolving power of $R\approx1600$. The resulting spectra cover a large wavelength range 3250--6215\,{\AA}, thus covering all of the hydrogen Balmer series lines from H$\beta$ to the Balmer jump. To achieve the highest possible $S/N$ -- as is required for accurate measurements of stellar magnetic fields -- we used the non-standard, 200kHz, low, 1$\times$1, readout mode. Further, to minimize the cross-talk effect, for each observation, we usually took four continuous series of two exposures at the position angles of the quarter-wave phase retarder plate $+45\degree$ and $-45\degree$ \citep[e.g.,][and references therein]{2002A&A...389..191B, 2004A&A...415..661H, 2004A&A...415..685H}. Wavelength calibration was made by arc lamp data taken at one retarder waveplate position, $+45\degree$ or $-45\degree$. The extraction of the parallel and perpendicular beams on the FORS\,2 raw data was carried out using a pipeline written in the MIDAS environment.

The mean longitudinal magnetic field $\left< B_{\rm z}\right>$ is the average over the stellar hemisphere visible at the time of observation of the component of the magnetic field parallel to the line of sight, weighted by the local emergent spectral line intensity. Its determination is based on the use of the equation

\begin{equation} 
\frac{V}{I} = -\frac{g_{\rm eff} e \lambda^2}{4\pi{}m_ec^2}\ \frac{1}{I}\ 
\frac{{\rm d}I}{{\rm d}\lambda} \left<B_{\rm z}\right>, 
\label{eqn:one} 
\end{equation} 

\noindent 
where $V$ is the Stokes parameter that measures the circular polarization, $I$ is the intensity in the unpolarized spectrum, $g_{\rm eff}$ is the effective Land\'e factor, $e$ is the electron charge, $\lambda$ is the wavelength, $m_e$ the electron mass, $c$ the speed of light, ${{\rm d}I/{\rm d}\lambda}$ is the derivative of Stokes $I$, and $\left<B_{\rm z}\right>$ is the mean longitudinal magnetic field. The effective Land\'e factor in our observations was set to 1.0 for the hydrogen lines and 1.1 elsewhere.

For each pair of subexposures obtained at the position angles of the retarder waveplate $+45\degree$ and $-45\degree$, $V/I$ was calculated using

\begin{equation}
\frac{V}{I} =
\frac{1}{2} \left\{ \left( \frac{f^{\rm o} - f^{\rm e}}{f^{\rm o} + f^{\rm e}} \right)_{\alpha=-45^{\circ}}
- \left( \frac{f^{\rm o} - f^{\rm e}}{f^{\rm o} + f^{\rm e}} \right)_{\alpha=+45^{\circ}} \right\},
\label{eqn:two}  
\end{equation}

\noindent
where $\alpha$ denotes the position angle of the retarder waveplate and $f^{\rm o}$ and $f^{\rm e}$ are ordinary and extraordinary beams, respectively. Stokes~$I$ values were obtained from the sum of the ordinary and extraordinary beams, which are recorded simultaneously by the detector and which were extracted using ESO MIDAS scripts. We sum up all corresponding $f^{e/o}_{\alpha=\pm45^{\circ}}$ to determine the longitudinal magnetic field from the full data set, after having ensured that the individual pairs lead to consistent results. To derive $\left<B_z \right>$, a least-squares technique was used to minimize the expression

\begin{equation}
\chi^2 = \sum_i \frac{(y_i - \left<B_z \right> x_i - b)^2}{\sigma_i^2}
\label{eqn:three}  
\end{equation}

\noindent 
where, for each spectral point $i$, $y_i = (V/I)_i$, $x_i = -\frac{g_{\rm eff} e \lambda_i^2}{4\pi{}m_ec^2}\ (1/I\ \times\ {\rm d}I/{\rm d}\lambda)_i$, and $b$ is a constant term that, assuming that Equation~\ref{eqn:one} is correct, approximates the fraction of instrumental polarization not removed after the application of Equation~\ref{eqn:two} to the observations. In all our measurements the fraction of instrumental polarization is well below 0.01\%.

Assuming that the only source of uncertainty in our field measurements is from the photon counting statistics of the observations, the longitudinal field uncertainty can be obtained from the formal uncertainty of the linear regression. Additionally, we have implemented a bootstrapping algorithm that determines the error of the magnetic field independently, without the need for assumptions on the statistics \citep{2010MNRAS.405L..46R,2014A&A...570A..88S}. The final magnetic field values and errors we present in this article are from the bootstrapping technique and agree very well with the values and errors determined with classical error propagation. For each spectral point $i$, the derivative of Stokes~$I$ with respect to the wavelength was evaluated following

\begin{equation}
  \left( \frac{dI}{d\lambda} \right)_{\lambda=\lambda_i} = \frac{N_{i+1}-N_{i-1}}{\lambda_{i+1}-\lambda_{i-1}},
\end{equation}

\noindent
where $N_i$ is the photon count at wavelength $\lambda_i$. 

The longitudinal magnetic field was measured in two ways: using only the absorption hydrogen Balmer lines ($\left<B_{\rm z}\right>_{\rm hyd}$) and using the entire spectrum including all available spectral lines ($\left<B_{\rm z}\right>_{\rm all}$). The results are provided in the final three columns of Table~\ref{tab:obs}, with the quoted errors being 1$\sigma$ uncertainties and with the final column giving measurements using the null spectrum for the whole spectral region, i.e. the spectrum computed by co-adding individual polarisation sub-exposures with signs changed in such a way as to cause the real polarisation signal to cancel out. The results for one of the FORS\,2 measurements are illustrated in Fig.~\ref{fig:MFs}.

\subsection{TESS Photometry} \label{tessphot}
TESS began its nearly all-sky survey in 2018, using four cameras with a combined field of view of 24$^{\circ}$ $\times$ 96$^{\circ}$. Each TESS field is observed for $\sim$27.4 days. The TESS bandpass extends over the range 6000--10500\,{\AA}, which includes the conventional $I$ and $Z$ and portions of the $V$, $R$, and $Y$ bands. Typical photometric precision is roughly 60 parts-per-million per hour. TESS observed Tr16-26 nearly continuously for $\sim$two months (between 2019 March 26 and 2019 May 21; sectors 10 and 11 with 30 minute cadence), followed by a $\sim$2 year gap, and then again for $\sim$two months (between 2021 March 7 and 2021 April 28; sectors 36 and 37 with 10 minute cadence.)

The full frame images (FFIs) were used to extract lightcurves ($lc$) for Tr16-26 using simple aperture photometry with a pixel mask tailored for each sector, which was necessary to avoid blending from nearby bright sources. A principal component analysis (PCA) algorithm was then used to remove systematic trends. The \textsc{lightkurve} \citep{Lightkurve2018} and \textsc{TESScut} \citep{Brasseur2019} Python packages were used for these steps. After removing outliers, a total of 9374 observations (1205 and 1241 in sectors 10 and 11, and 3467 and 3461 in sectors 36 and 37) were used the analysis presented in this paper.

Due to the combination of the crowded field that Tr16-26 resides in and the size of the TESS pixels ($\approx23\arcsec$ per side), blending is unavoidable. Of particular concern is the star Tr16-25 ($V=11.6$), which is separated on the sky from Tr16-26 by just 19$\farcs$7 and slightly brighter than Tr16-26. As a first step to addressing this potential issue, a pixel-level analysis of the FFIs was performed to localize the origin of various signals in the field. This analysis revealed some variable sources in the vicinity of Tr16-26, but confirmed that the signal of interest -- rotational modulation -- originates on target and that neighboring sources do not contribute any significant signals to the extracted light curve. However, as will be discussed subsequently, the amplitude of rotational modulation of Tr16-26 is in fact diluted by Tr16-25. 

For comparison to known CM stars, the same process above was also applied to the TESS $lc$ of {\sigorie}, HD\,23478, and HD\,345439. In the case of HD\,345439, which was pre-selected for higher cadence TESS observations, the 2-minute cadence $lc$ was downloaded from MAST and the PDCSAP flux was used (identical signals are found in the 2-minute SAP flux, and also in $lc$ extracted from the FFIs.)

\begin{figure*}
\includegraphics[width=\textwidth]{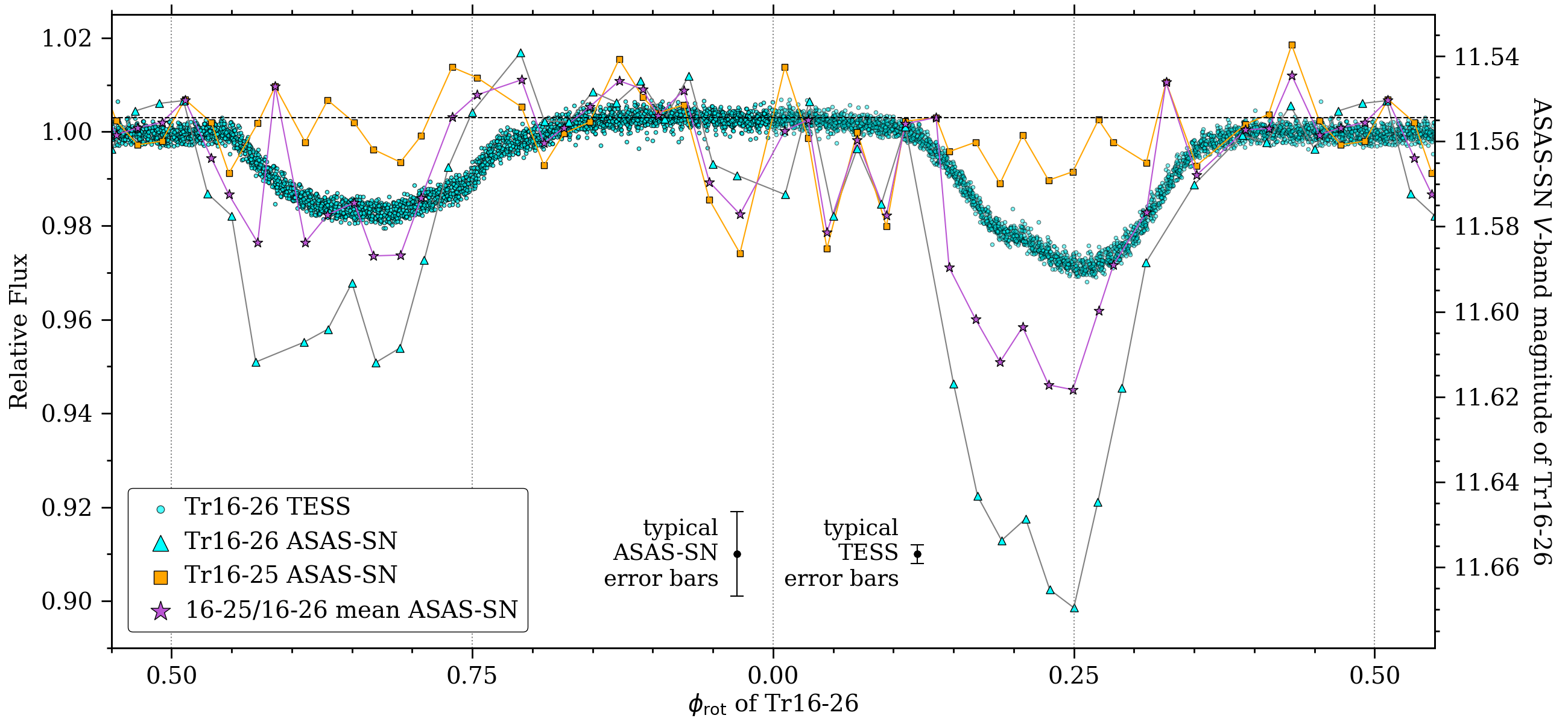}
\caption{The TESS and ASAS-SN $lc$ of Tr16-26. To demonstrate the fact that the TESS $lc$ is contaminated by stars falling on the same or adjacent pixels, the ASAS-SN $lc$ of the close neighbor star Tr16-25 is also shown, along with the mean ASAS-SN $lc$ of Tr16-26 and Tr16-25. All of the data have been phased by the 0.9718115\,day rotation period of Tr16-26, and the ASAS-SN data have been binned in increments of $\phi_{\rm rot}=0.02$. The horizontal dashed line indicates the approximate out-of-eclipse $V$ magnitude of Tr16-26, and vertical dotted lines indicate $\phi_{\rm rot}$ increments of 0.25. \label{fig:lc}}
\end{figure*}

\begin{figure}
\includegraphics[width=\columnwidth]{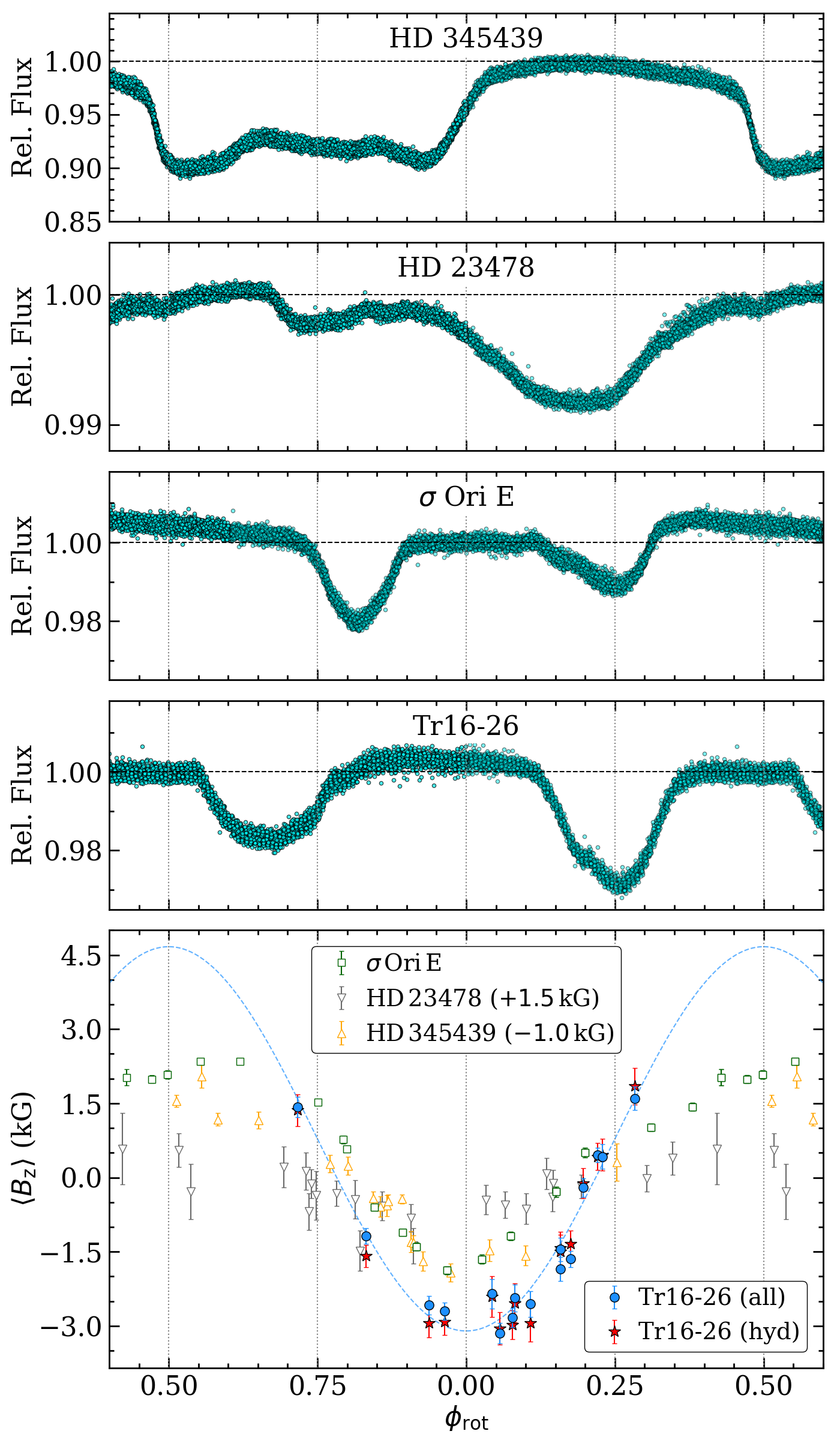}
\caption{\emph{Upper panels}: Rotation-phased TESS lightcurves for HD\,345439, HD\,23478, {\sigorie}, and Tr16-26. For {\sigorie}, HD\,23478, and HD\,345439 the data are phased via the periods reported by \citet{2010ApJ...714L.318T, 2015MNRAS.451.1928S}, and \citet{2015ApJ...811L..26W}, but with the phases adjusted such that {\bzmin} occurs at $\phi_{\rm rot}=0$. Horizontal dashed lines indicate approximately relative flux of unity, and vertical dotted lines indicate $\phi_{\rm rot}$ increments of 0.25. \emph{Bottom panel}: The {\bz} measurements of Tr16-26 (filled symbols) phased by the rotation period determined from the TESS lightcurve, with the dashed curve being the best fitting sinusoid to the $\left<B_{\rm z}\right>_{\rm all}$ measurements. Unfilled symbols are {\bz} measurements for {\sigorie} \citep{2012MNRAS.419..959O}, HD\,23478 \citep{2015MNRAS.451.1928S}, and HD\,345439 \citep{2017MNRAS.467L..81H}. \label{fig:bzlc}}
\end{figure}

\subsection{ASAS-SN Photometry} \label{asas}
We also downloaded ASAS-SN $V$-band lightcurves for both Tr16-26 and the close neighbor star Tr16-25. The ASAS-SN survey provides full-sky, time series photometry obtained by numerous 14-inch telescopes. Given the size of the pixels in the associated CCDs (7$\farcs$8), the ASAS-SN data can provide a sanity check on the TESS data in cases of crowded fields like that of Tr16-26.

\section{Lightcurve and Rotation Period} \label{prot}
Tr16-26 has previously been classified as an eclipsing binary (EB) based on the ASAS-SN $lc$, from which a 0.971844\,day period was estimated based on two unequal eclipse-like events that repeat each period. Despite the similarity to EBs, our spectra of Tr16-26 show no evidence of a second set of absorption lines nor of radial velocity variability of the magnetic star. Instead, we argue that the eclipses are caused by occultations of the stellar disk by magnetically confined clouds of circumstellar gas. This is a photometric hallmark of rapidly rotating CM stars \citep[e.g.,][]{2005ApJ...630L..81T}.

Figure~\ref{fig:lc} displays the TESS $lc$ of Tr16-26 along with the ASAS-SN $lc$ of both Tr16-26 and the near neighbor Tr16-25. The latter star was classified as a double-lined spectroscopic binary (SB2) by H18, but the ASAS-SN $lc$ is essentially flat, showing little to no evidence of variability. On the other hand, the Tr16-26 eclipses are blatant in both the ASAS-SN and TESS $lc$. Although this confirms that the photometric variability arises from Tr16-26 rather than any nearby stars, the mismatched eclipse depths between the two data sets also indicates that the TESS $lc$ is in fact contaminated by Tr16-25 and perhaps also by other neighboring stars since there is still a mismatch between eclipse depths if we consider the simple mean of the ASAS-SN $lc$ for Tr16-26 and Tr16-25 (star points and purple lines in Figure~\ref{fig:lc}).  Since the precise eclipse depths are not critical for a period search in this case, we proceeded with analysis of the TESS $lc$ under the assumption that no other variable sources aside from Tr16-26 contribute to the $P_{\rm rot}\approx1$\,day rotational modulation. As will be shown, the safety of this assumption is supported by the spectropolarimetry.

To refine the rotation period of Tr16-26 estimated by ASAS-SN, we analyzed the combined TESS $lc$ consisting of all four sectors of data. Although the photometric variability is far from sinusoidal, a Fourier-like analysis is still a reasonable method to determine the period of the variations. The time series analysis software package \textsc{Period04} \citep{Lenz2005} was used, and a rotational period of $P_{\rm rot} = 0.9718115 \pm 5.7 \times 10^{-6}$ days (or $f_{\rm rot} = 1.0290062 \pm 6.1 \times 10^{-6}$ days$^{-1}$) was found. Note that because the rotational photometric signal is double waved (\textit{i.e.} there are two dips of unequal depth which are both seen in one rotational cycle), the strongest signal in a frequency spectrum is at half of the stellar rotation period. Virtually identical periods are found using other methods, for example the Box Least Squares (BLS) algorithm, which is commonly used for eclipsing binaries and exoplanet transits. Instead of fitting sinusoids (as Period04 does), BLS fits an eclipse-like box to the data. Regardless of the analysis method, inspection of the phased TESS photometry leaves no doubt that the correct value for $P_{\rm rot}$ has been recovered.

The TESS $lc$ of Tr16-26 is compared to those of the previously known CM stars HD\,345439, HD\,23478, and {\sigorie} in the upper panels of Figure~\ref{fig:bzlc}. Whereas the HD\,345439 and HD\,23478 lightcurves are highly irregular, with wide ($\sim$half of a rotation period) eclipse-like events exhibiting multiple minima, the {\sigorie} and Tr16-26 lightcurves are relatively orderly and remarkably similar to one another. When the photometry is phased such that the minimum longitudinal magnetic field strength occurs at $\phi_{\rm rot}=0$, the minima of the more irregular eclipses coincide almost exactly with $\phi_{\rm rot}=0.25$. The out-of-eclipse behavior of {\sigorie} and Tr16-26 is also similar, with the baseline fluxes being slightly higher following the shallower eclipse than they are after the deeper eclipse. Despite the successes of the RRM model in duplicating many observational aspects of CM stars, attempts to reproduce the lightcurve of {\sigorie} have thus far been only partially successful \citep[see Figure 8 of][]{2015MNRAS.451.2015O}.

\section{Magnetic Field} \label{bfield}
Filled symbols in the bottom panel of Figure~\ref{fig:bzlc} are the results of phasing our Tr16-26 {\bz} measurements to the period obtained from the TESS $lc$. For comparison, {\bz} measurements are also shown for {\sigorie} \citep{2012MNRAS.419..959O}, HD\,23478 \citep{2015MNRAS.451.1928S}, and HD\,345439 \citep{2017MNRAS.467L..81H}, with the measurements of the latter two stars having been shifted vertically for clarity by $+1.5$\,kG and $-1.0$\,kG, respectively. 

Several things are immediately clear from this comparison. First and foremost is the fact that our phase coverage of Tr16-26 is marginal at best, covering only slightly more than half of a rotation period. Second, it is likely that the magnetic field of Tr16-26 is stronger than those of HD\,23478 and HD\,345439, and perhaps also of {\sigorie}. Third, whereas the {\bz} variations of Tr16-26 and {\sigorie} appear to be mostly sinusoidal around {\bzmin}, they are decidedly non-sinusoidal around {\bzmax} in the case of {\sigorie}. Thus, although it is obvious that Tr16-26 is highly magnetized, it is not clear from our data whether or not the magnetic field can be treated as a simple dipole. Further, the fact that the eclipses are not neatly separated by $\phi_{\rm rot}=0.5$ for {\sigorie} nor Tr16-26 hints that, as in the case of {\sigorie} \citep{2015MNRAS.451.2015O}, the magnetic field of Tr16-26 may also have a significant quadrupolar component. 

The dashed blue line in the bottom panel of Figure~\ref{fig:bzlc} is the sine curve that best fits our $\left<B_{\rm z}\right>_{\rm all}$ measurements. If we were to assume that the Tr16-26 magnetic field is purely dipolar and that the blue curve is an accurate extrapolation of the {\bz} variation, the approximation of \citet{1950MNRAS.110..395S} and \citet{1967ApJ...150..547P} could be used to estimate the dipole field strength. Combining the sine curve parameters with an appropriate limb darkening coefficient \citep[$\mu\sim0.35$;][]{2016MNRAS.456.1294R} would indicate $B_{\rm d}\approx19$\,kG, making Tr16-26 the second most magnetic CM star known behind only ALS\,2394 \citep[a.k.a. CPD\,$-62^{\circ}\,2124$, $B_{\rm d}\sim21$\,kG;][]{2017MNRAS.472..400H}. Due to the poor phase coverage however, we can only establish a lower limit on $B_{\rm d}$. Assuming {\bz} turns over around $\phi_{\rm rot}=\pm0.25$ (similar to the behavior of {\sigorie}), then we would have {\bzmin}$\approx-3.1$\,kG and {\bzmax}$\approx1.6$\,kG. These values indicate $B_{\rm d}>11.8$\,kG, such that Tr16-26 is undoubtedly among the most magnetic OB stars known.

The {\bz} curve of Tr16-26 is also useful for establishing a meaningful rotational/magnetic ephemeris. From the sine curve fit in the bottom panel of Figure~\ref{fig:bzlc}, we find that $\rm JD=2457000.1770$ is an epoch corresponding to {\bzmin}. Use of the $\left<B_{\rm z}\right>_{\rm hyd}$ measurements yields a virtually identical result.

\section{Spectral Classification and Stellar Parameters} \label{spectype}

\begin{figure*}
\includegraphics[width=450pt]{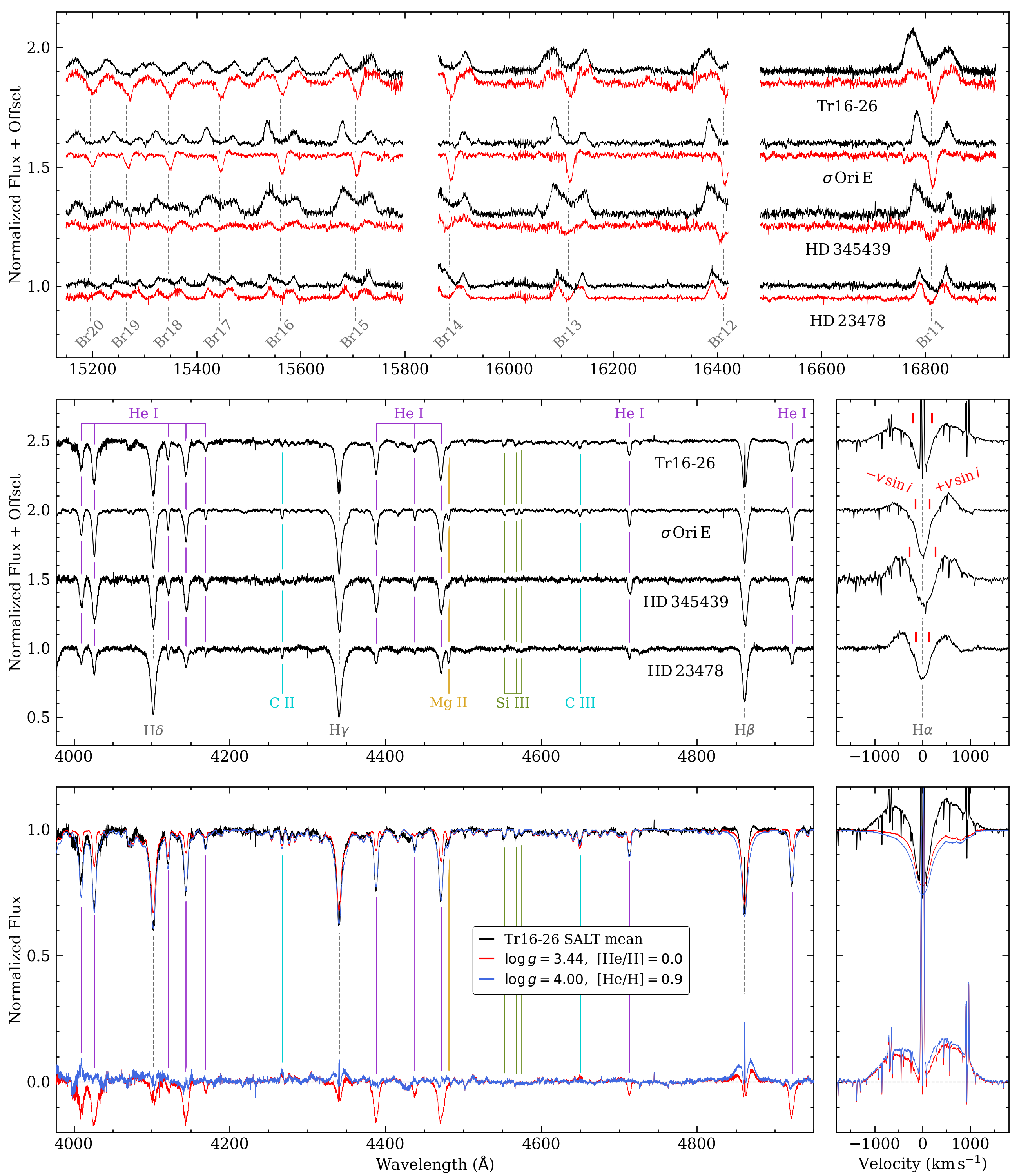}
\caption{\emph{Top}: Comparison of the minimum (red) and maximum (black) emission APOGEE spectra of the CM stars Tr16-26, {\sigorie}, HD\,345439, and HD\,23478. The hydrogen Brackett series lines are labeled along the bottom. \emph{Middle}: Comparison of the average SALT/HRS spectrum of Tr16-26 to ARC/ARCES spectra of the comparison sample. Most of the strong atomic lines are labeled. Red ticks in the H$\alpha$ panel indicate $\pm\,v \sin i$ as reported by H18 for Tr16-26, and as reported by \citet{2019MNRAS.485.1508S} for {\sigorie}, HD\,345439, and HD\,23478. Note that the narrow emission features in the Tr16-26 spectra are nebular rather than circumstellar in nature, and also that the HD\,345439 spectrum has been smoothed due to relatively low $S/N$. \emph{Bottom}: Comparison of the average SALT/HRS spectrum of Tr16-26 to synthetic spectra with $T_{\rm eff}=24,800$\,K and $v\,\sin\,i=195$\,km\,s$^{-1}$. The red synthetic spectrum has the $\log g=3.44$ reported by H18 and Solar helium abundance. The light blue synthetic spectrum has $\log g=4.00$ and helium abundance enhanced by 0.9 dex. Synthetic minus observed residuals are shown along the bottom.\label{fig:apogeespec}}
\end{figure*}

The upper panel of Figure~\ref{fig:apogeespec} compares APOGEE spectra of Tr16-26 to the previously known CM stars HD\,345439, {\sigorie}, and HD\,23478. For each star, red lines represent the minimum available emission spectra and black lines pertain to the maximum available emission spectrum. The transformation from pure emission to almost pure absorption in the {\sigorie} spectra is remarkable, especially considering that the cycle repeats every $1.19$\,d. The temporal behavior of Tr16-26 and HD\,345439 is similar, albeit with the minimum (available) emission spectra still exhibiting considerable emission above continuum level.

As for HD\,23478, the relative constancy of the H$\alpha$ emission has been noted by \citet{2020MNRAS.499.5379S} and is likely a result of its low obliquity angle ($\beta=4^{\circ}$), which serves to diminish the amplitude of any rotational modulation. The APOGEE spectra are no different, with the Brackett series lines maintaining a similar, pure emission spectrum throughout the 26 $H$-band observations. Figure~\ref{fig:apogeespec} therefore simply shows two of the most different HD\,23478 spectra. 

In addition to closely resembling other CM stars in the $H$-band, Tr16-26 is a near twin of {\sigorie} and HD\,345439 in the optical. This is demonstrated in the middle panel of Figure~\ref{fig:apogeespec}, which shows portions of the average SALT/HRS spectrum of Tr16-26 above the ARC/ARCES spectra of HD\,345439, {\sigorie} and HD\,23478. The corresponding H$\alpha$ line profiles are shown at right on a velocity scale, emphasizing the fact that most of the emission is occurring beyond $v \sin i$. 

Whereas HD\,23478 appears to be more or less He-normal, the other three stars are blatantly He-enhanced and also slightly hotter than HD\,23478 based on clear detection of lines from Si~{\sc iii}, C~{\sc iii}, and O~{\sc ii} and also on the weakness of lines from lower ionization stages (e.g., the Si~{\sc ii} 4128/4130 {\AA} lines, which are clearly present for HD\,23478.) Based on the strength of the Si~{\sc iii}, C~{\sc iii}, and O~{\sc ii} of Tr16-26 relative to {\sigorie}, and accounting for the more rapid rotation of Tr16-26, we see no reason to adjust the B1.5\,V literature spectral type beyond making it a more specific B1.5\,Vpe\,He-strong.

To investigate the discrepancy between the parameters reported by H18 and the Gaia eDR3 distance (see the concluding paragraphs of Section~\ref{intro}), we used the \textsc{SYNSPEC} \citep{2017arXiv170601859H} program fueled with Kurucz\footnote{1995 Atomic Line Data (R.L. Kurucz and B. Bell) Kurucz CD-ROM No. 23. Cambridge, Mass.: Smithsonian Astrophysical Observatory.} LTE model atmospheres to create a synthetic spectrum with the parameters reported by H18. In the bottom row of Figure~\ref{fig:apogeespec}, the black line is the average SALT/HRS spectrum and the red line is the aforementioned synthetic spectrum. Observed minus synthetic residuals are shown along the bottom, and from those it is immediately confirmed that Tr16-26 is He-enhanced. More importantly however, the $\log g=3.44$ residuals around the Balmer series lines indicate an additional hydrogen absorption component rather than the emission component that is expected given that three out of the four SALT/HRS spectra were taken at out-of-eclipse rotational phases (where there shouldn't be much additional absorption caused by the circumstellar gas crossing the observer line of sight.) Otherwise, the H18 parameters provide a satisfactory fit to most of the metal absorption lines, and we see no reason to revise $T_{\rm eff}$ nor $v \sin i$. 

A $\log g$ much larger than 3.44 is needed to put Tr16-26 at a distance consistent with the Gaia distance, and to demonstrate the associated effects on the residuals, the blue line in the bottom row of Figure~\ref{fig:apogeespec} shows a synthetic spectrum that is identical to the red one except for having $\log g=4.00$ and [He/H]$=+0.9$. This results in a vastly improved match to the observed spectrum, particularly in terms of the the He~{\sc i} and Balmer series lines. One exception to that statement is the C~{\sc ii}~4267\,{\AA} doublet, which is well known to be poorly fit by LTE models \citep[e.g.,][]{1988A&A...202..153E}.

\begin{table}
\caption{Measured and adopted parameters of Tr16-26. \label{tab:parms}}
\begin{tabular}{lrr}
\hline
Parameter  & Value & Uncertainty \\ 
\hline
Gaia $\mu$ (mas\,yr$^{-1}$)                                            & 7.269                     & 0.018                    \\ 
Gaia $\pi$ (mas)                                                       & 0.4038                    & 0.0143                   \\ 
Gaia distance (pc)                                                     & 2261                      & 215                      \\ 
$P_{\rm rot}$ (d)                                                      & 0.9718115                 & 0.0000057                \\ 
$\langle B_{\rm z}\rangle_{\rm min}$ (kG)                              & $-3.097$                  & 0.120                    \\ 
$T_{\langle B_{\rm z}\rangle_{\rm min}}$ (JD$-2.4$e6)                  & 57000.1770                & 0.0010                   \\ 
$m_{V}$ (mag)                                                          & 11.55                     & 0.02                     \\ 
$T_{\rm eff}$ (K)                                                      & 24800                     & 1000                     \\ 
$\log g$                                                               & 4.00                      & 0.20                     \\ 
$v\,$sin$\,i$ (km$\,s^{-1}$)                                           & 195                       & 20                       \\ 
$R_{\rm eq}$ ($R_{\odot}$)                                             & 5.13                      & 0.42                     \\ 
$M_{*}$ ($M_{\odot}$)                                                  & 9.71                      & 1.53                     \\ 
$\log L_{*} / L_{\odot}$                                               & 3.95                      & 0.14                     \\ 
$M_{V}$ (mag)                                                          & $-$2.68                   & 0.25                     \\ 
$A_{V}$ (mag)                                                          & 2.46                      & 0.32                     \\ 
$i_{\rm rot}$ ($^{\circ}$)                                             & 46.9                      & 8.1                      \\ 
$v_{\rm eq}$ (km$\,$s$^{-1}$)                                          & 267                       & 22                       \\ 
$v_{\rm orb}$ (km$\,$s$^{-1}$)                                         & 599                       & 27                       \\ 
$W=v_{\rm eq}/v_{\rm orb}$                                             & 0.446                     & 0.042                    \\ 
$R_{\rm K}$ ($R_{\rm eq}$)                                             & 1.71                      & 0.05                     \\ 
\hline
\textbf{H$\alpha$ Max. Emission ($\phi_{\rm rot}=0.45$)} & & \\ 
$|W_{\rm H\alpha,max}|$ (nm)                                           & 0.473                     & 0.053                    \\ 
$r_{\rm max}$ ($R_{\rm eq}$)                                           & 2.8                       & 0.4                      \\ 
$r_{\rm out}$ ($R_{\rm eq}$)                                           & 7.2                       & 0.9                      \\ 
V/R ($\phi_{\rm rot}=0.45$)                                            & 1.35                      & 0.03                     \\ 
$r_{\rm max}/R_{\rm K}$                                                & 1.65                      & 0.23                     \\ 
\hline
\end{tabular}
\end{table}

Our measured and/or adopted observational parameters for Tr16-26 are given in Table~\ref{tab:parms}, beginning with the total proper motion ($\mu$), parallax ($\pi$), and distance from Gaia eDR3. The quoted error on the distance to Tr16-26 is simply the difference between $1/\pi$ and the 2261\,pc of \citet{2021AJ....161..147B}. The $\mu=7.269\pm0.018$\,mas\,yr$^{-1}$ is well within the error bars of the average for Trumpler 16 members \citep[$\overline{\mu}_{\rm Tr16}=7.417\pm0.327$\,mas\,yr$^{-1}$;][]{2021ApJ...914...18S}, thus solidifying Tr16-26 as a main sequence member of Trumpler 16 rather than a background subgiant/giant. The quoted apparent $V$-band magnitude ($m_{V}=11.55\pm0.02$) was estimated from out-of-eclipse phases in the ASAS-SN $lc$ (indicated by the dashed horizontal line in Figure~\ref{fig:lc}). After comparison of the average SALT/HRS spectrum to synthetic spectra with a range of parameters close to the adopted values, we revised the uncertainties on $T_{\rm eff}$ (from the $\pm2000$\,K reported by H18 to $\pm1000$\,K), $\log g$ (from $\pm0.22$ dex to $\pm0.20$ dex), and $v \sin i$ (from $\pm37$\,km\,s$^{-1}$ to $\pm20$\,km\,s$^{-1}$.)

We also obtained estimates of the luminosity ($\log L_{*}/L_{\odot}=3.95\pm0.14$) and absolute $V$-band magnitude ($M_{V}=-2.68\pm0.25$) of Tr16-26 by comparing our $T_{\rm eff}$ and $\log g$ estimates to an interpolated version of the grid\footnote{\url{http://www.pas.rochester.edu/~emamajek/EEM_dwarf_UBVIJHK_colors_Teff.txt}} of stellar parameters associated with \citet{2013ApJS..208....9P}. The quoted errors represents the range of possibilities given the uncertainties on $T_{\rm eff}$ and $\log g$. We then proceeded to estimate a classical radius of $R_{*}=5.13\pm0.42\,R_{\odot}$ via the relation $R_{*}/R_{\odot}=\sqrt{(L_{*}/L_{\odot})/(T_{\rm eff}/T_{\rm eff,\odot})^4}$, a spectroscopic mass of $M_{*}=9.71\pm1.53\,M_{\odot}$ via the relation $M_{*}/M_{\odot}=g(R_{*}/R_{\odot})^{2}/G$, and a $V$-band extinction of $A_{V}=2.46\pm0.32$ via the distance modulus. The latter value is quite high, but not at all unreasonable given the wide range of extinction and occasionally very high extinction of members of the Trumpler 14, 15, and 16 \citep[often around 2.5\,mag and up to $\sim3$\,mag;][]{2021ApJ...914...18S}.

Although Tr16-26 rotates sufficiently rapidly that it is unlikely to be spherical, some further estimates can be made under the assumption that $R_{*}$ is the same as the equatorial radius, $R_{\rm eq}$. In particular, the rotational inclination angle ($i_{\rm rot}$) and rotational velocity at the equator ($v_{\rm eq}$) can be determined via sin\,$i=(P_{\rm rot}\,v\,$sin$\,i)/(2\pi\,R_{\rm eq}$), and for Tr16-26, we find $i_{\rm rot}=46.9\pm8.1^{\circ}$ and $v_{\rm eq}=267\pm22$\,km\,s$^{-1}$. Also of interest are $v_{\rm orb}$, which is the velocity required for circular, Keplerian orbit near the equatorial surface ($v_{\rm orb}=\sqrt{G\,M_{*}/R_{\rm eq}}$; also commonly referred to as the critical velocity), and $W$, a dimensionless rotation parameter ($W=v_{\rm eq}$/$v_{\rm orb}$) that is zero in the case of no rotation, one in the case of critical rotation \citep{2013A&ARv..21...69R}, and $0.446\pm0.042$ in the case of Tr16-26.

Given the estimate of $W$, the Kepler co-rotation radius at the equator can be estimated according to $R_{\rm K}=W^{-2/3}R_{\rm eq}$. For Tr16-26, we find $R_{\rm K}=1.71\pm0.05\,R_{\rm eq}$, which is among the smallest $R_{\rm K}$ of the CM stars studied by \citet{2020MNRAS.499.5379S}. Although a robust determination of the Alfv\'{e}n radius ($R_{\rm A}$) is impossible with the available data, improved {\bz} phase coverage will undoubtedly confirm that $R_{\rm A}$ greatly exceeds $R_{\rm K}$, thus enabling existence of a CM.

\begin{figure*}
\includegraphics[width=\textwidth]{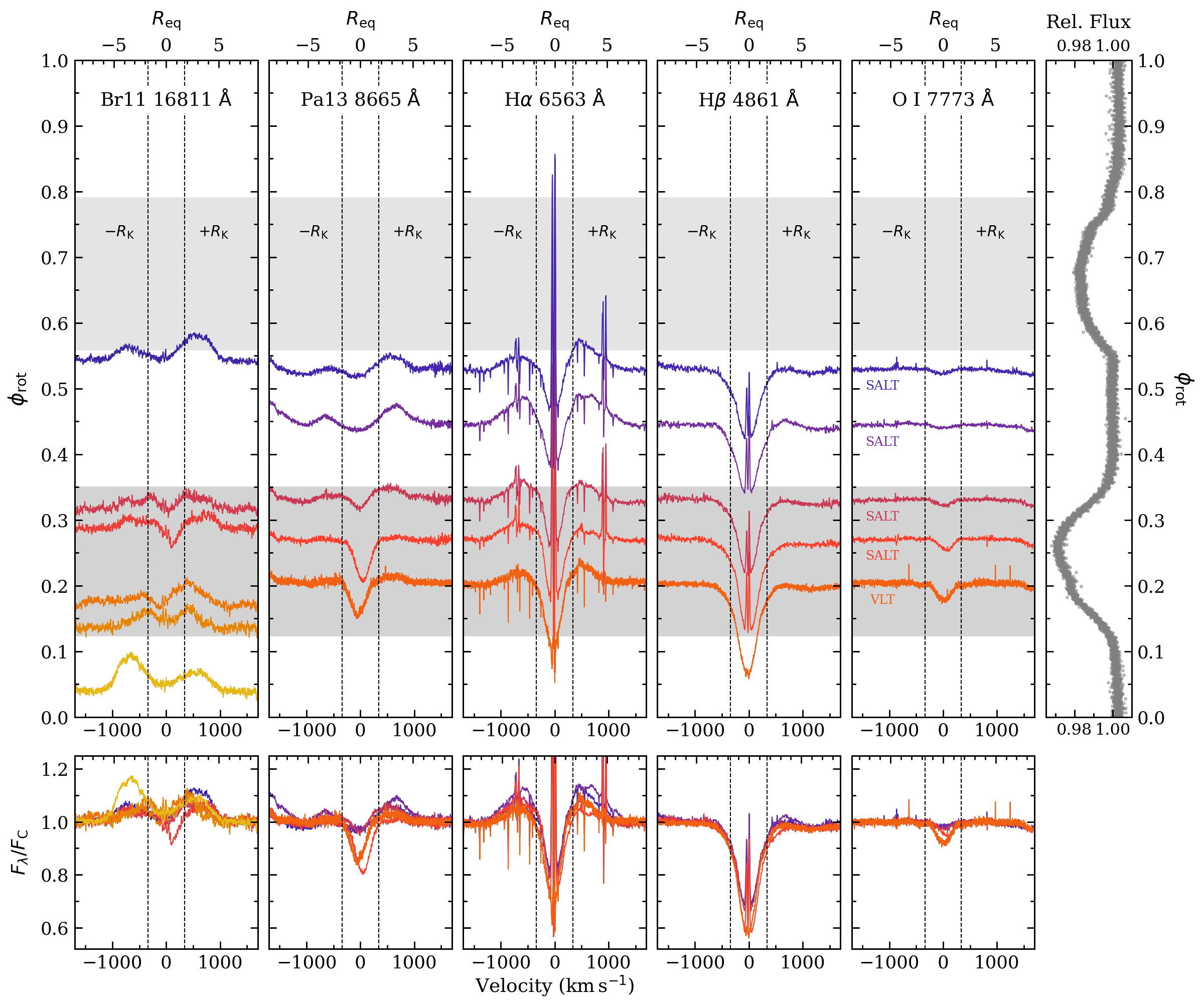}
\caption{Spectroscopic variability of Tr16-26 as a function of rotational phase. Individual spectral lines are shown on a velocity scale in the the left five columns, and for context, the TESS lightcurve is shown in the far right panel. In the upper panels of the spectral line columns, colors and vertical spacing of the spectra are proportional to $\phi_{\rm rot}$. The leftmost column shows the strongest Brackett series line (Br11; 11--4) covered by the APOGEE spectra. Otherwise, the SALT/HRS and VLT/UVES spectra are displayed (as indicated in the O~{\sc i} panel). Grey shading indicates phases corresponding to photometric eclipses. The top x-axes are in units of equatorial radius ($R_{\rm eq}$), and vertical dashed lines correspond to the Kepler co-rotation radius ($\pm R_{\rm K}$). \label{fig:specvar}}
\end{figure*}

\section{Spectroscopic Variability and H$\alpha$ Emission Parameters} \label{specvar}
 Despite the sparse $\phi_{\rm rot}$ coverage of our spectroscopic data set, hints of the expected phase locked variability are seen. This is demonstrated in Figure~\ref{fig:specvar}, which shows portions of the APOGEE, VLT/UVES, and SALT/HRS spectra alongside the TESS lightcurve.

At left is the Br11 line captured by the APOGEE spectra, exemplifying the emission dominated and highly variable Brackett series lines. With an absolute equivalent width of $|W_{\rm Br11}|\approx0.9$\,nm, the $\phi_{\rm rot}=0.05$ spectrum represents the maximum emission APOGEE spectrum. Correcting for underlying photospheric absorption would only increase this estimate. The ratio of the violet ($V$) and red ($R$) peak heights is also at a maximum around $\phi_{\rm rot}=0.05$ ($V$/$R\sim1.06$), and the velocity separation of $\sim1225$\,km\,s$^{-1}$ is seconded only by the $\sim1300$\,km\,s$^{-1}$ peak separation of the $\phi_{\rm rot}=0.55$ spectrum. Given the phase difference of almost 0.5 between these spectra and the fact that the line profiles are essentially mirror images, the $\phi_{\rm rot}=0.05$ and 0.55 spectra represent our best evidence not only of phase-locked variability of the hydrogen emission, but of double-waved spectroscopic variability \citep[expected when $i_{\rm rot}$ and $\beta$ are both high][]{2020MNRAS.499.5379S}. 

As for the APOGEE spectra in rotation phases $\phi_{\rm rot}\approx0.1$--0.35, most of which were observed during the shallow, irregular eclipse, the situation is more chaotic. The overall emission strength decreases, and the minimum Br11 emission ($|W_{\rm Br11}|\approx0.2$\,nm) is observed in the $\phi_{\rm rot}=0.29$ spectrum. In addition to the relative weak emission at $\phi_{\rm rot}=0.29$, the Brackett series lines exhibit high velocity, narrow absorption (or `shell') components that are reminiscent of the behavior of {\sigorie} \citep{1976A&A....52..303G}.

The Pa13, H$\alpha$, and H$\beta$ panels of Figure~\ref{fig:specvar} exhibit similar behavior. For example, whereas the Pa13 line takes the form of a blueshifted, mostly absorption profile in the $\phi_{\rm rot}=0.21$ spectrum, the feature has migrated to high velocities in the $\phi_{\rm rot}=0.28$ spectrum. As with the Brackett series, the optical emission is strongest during the out-of-eclipse phases, when both accumulations of rigidly rotating gas are visible to the observer. The Paschen series lines are particularly remarkable in terms of transitioning from almost pure absorption during the shallow eclipse to pure emission between eclipses. 

All of the helium and metal lines are at least mildly variable in our spectra, with the overall most variable non-hydrogen feature being the blend of the O~{\sc i} triplet between 7771 and 7775\,{\AA}. Unlike the hydrogen lines, the apparent filling of the O~{\sc i} cores with emission during out-of-eclipse phases is not accompanied by wide, double peaked emission. 

As can be seen in Figure~\ref{fig:specvar}, the maximum H$\alpha$ emission of our dataset was the $\phi_{\rm rot}=0.45$ observation. In order to put Tr16-26 into context with the \citet{2020MNRAS.499.5379S}  sample of CM stars, we attempted to measure the equivalent width of the H$\alpha$ emission. For that purpose, our first step was removing the photospheric absorption component from the $\phi_{\rm rot}=0.45$ spectrum by dividing out the SYNSPEC model spectrum shown as a blue line in the lowermost panel of Figure~\ref{fig:apogeespec}. Next, regions coinciding with narrow nebular and telluric contaminants were interpolated over based on the flux levels of smooth adjacent regions. We then measured $|W_{\rm H\alpha,max}|$ the same way as \citet{2020MNRAS.499.5379S}, where the region of the H$\alpha$ between $\pm v \sin i$ is excluded due to concerns about added absorption from the circumstellar disk and about synthetic spectra failing to match the line core perfectly. Direct integration of the flux in the regions between $\pm v \sin i$ and the outer edges of the line profile and summing the two measurements gives a lower limit (due to sparse phase coverage) of $|W_{\rm H\alpha,max}|=0.473\pm0.053$\,nm, where the error is the average standard deviation of continuum regions adjacent to the line profile, multiplied by ten to be conservative given the telluric and nebular contamination of the observed line profile. 

Regardless of the details of how $|W_{\rm H\alpha,max}|$ is estimated, Tr16-26 is undoubtedly among the strongest emission producers of the known CM sample. The only other CM star with $|W_{\rm H\alpha,max}|>0.35$\,nm is HD\,345439, for which \citet{2020MNRAS.499.5379S} reported $|W_{\rm H\alpha,max}|=0.478\pm0.009$\,nm.

We also estimated the velocities of the H$\alpha$ emission peaks ($v_{\rm max}$) and outer edges ($v_{\rm out}$) relative to the line center. These are quantities of interest because, under the assumption of rigid rotation of the emitting gas, their velocities correspond directly to physical distances from the star referred to by \citet{2020MNRAS.499.5379S} as $r_{\rm max}$ and $r_{\rm out}$. The H$\alpha$ line profile of Tr16-26 around $\phi_{\rm rot}=0.45$ is sufficiently complicated and non-Gaussian that we did not attempt to model it. Rather, we estimated $v_{\rm out}$ by eye and $v_{\rm max}$ by interactively fitting a blend of two Gaussians to the line profile using the $splot$ program in IRAF.  

For $v_{\rm out}$, we find $-1300$\,km\,s$^{-1}$ and $+1500$\,km\,s$^{-1}$ and a ballpark uncertainty of $\pm50$\,km\,s$^{-1}$, confirming that the line profile is asymmetric. Following \citet{2020MNRAS.499.5379S}, we adopt the velocity corresponding to the stronger emission peak, which in this case means $v_{\rm out}\approx1500\pm50$\,km\,s$^{-1}$. Similarly, we find $v_{\rm max}\approx550\pm50$\,km\,s$^{-1}$. Dividing these velocities by $v \sin i$ yields $r_{\rm max}\approx2.8\pm0.4$\,$R_{\rm eq}$ and $r_{\rm out}\approx7.2\pm0.9$\,$R_{\rm eq}$. These values are similar to the $r_{\rm max}\approx3.0\pm0.1$\,$R_{\rm eq}$ and $r_{\rm out}\approx6.9\pm0.3$\,$R_{\rm eq}$ for {\sigorie}, and they imply that Tr16-26 has the third largest H$\alpha$-emitting volume of the known sample of CM stars behind HD\,164492C and ALS\,2394 \citep{2020MNRAS.499.5379S}. The maximum H$\alpha$ emission parameters as observed in the $\phi_{\rm rot}=0.45$ SALT/HRS spectrum are provided toward the bottom of Table~\ref{tab:parms}.

\section{Concluding Remarks}
Although stars like {\sigorie} continue to violate the expectation of OB stars lacking both built-in magnetic fields and the ability to generate them via convection, tremendous progress has been made over the last decade in identifying, characterizing, and classifying magnetic OB stars. The importance of expanding the known sample is clear, and to that end, we have confirmed the  B1.5\,V star Tr16-26 as not only a CM host, but as one of the most extreme examples (comparable to {\sigorie}, HR\,5907, HR\,7355, and HD\,345439.) Consistent with the expectations set by the known sample of CM host stars \citep{2019MNRAS.490..274S,2020MNRAS.499.5379S}, Tr16-26 is a rapid rotator with a strong magnetic field, and almost certainly a young star given the stellar parameters reported here and membership in the Trumpler 16 open cluster \citep[also home to $\eta$\,Car and numerous stars earlier than O3;][]{2008hsf2.book..138S}. 

Tr16-26 is the third CM star discovered via SDSS/APOGEE $H$-band spectroscopy along with HD\,23478 and HD\,345439. The fact that these three stars alone account for 14.3\% of the known CM sample solidifies the notion of multi-epoch, near-infrared spectroscopy as a powerful tool for detecting the wide hydrogen emission and associated rapid variability. It is likely that additional CM examples with weaker emission are lurking in the existing APOGEE data, and finding and confirming them represents ongoing work for us. We expect additional discoveries to be made during the upcoming fifth instalment of the SDSS survey, in which the fiber plug-plate system will be replaced by fiber positioning robots, thus vastly increasing the data volume of the survey.

\section*{Acknowledgements}

Funding for the Sloan Digital Sky Survey IV has been provided by the Alfred P. Sloan Foundation, the U.S. Department of Energy Office of Science, and the Participating Institutions. SDSS acknowledges support and resources from the Center for High-Performance Computing at the University of Utah. The SDSS web site is www.sdss.org.

SDSS is managed by the Astrophysical Research Consortium for the Participating Institutions of the SDSS Collaboration including the Brazilian Participation Group, the Carnegie Institution for Science, Carnegie Mellon University, the Chilean Participation Group, the French Participation Group, Harvard-Smithsonian Center for Astrophysics, Instituto de Astrof\'{i}sica de Canarias, The Johns Hopkins University, Kavli Institute for the Physics and Mathematics of the Universe (IPMU) / University of Tokyo, Lawrence Berkeley National Laboratory, Leibniz Institut f\"{u}r Astrophysik Potsdam (AIP), Max-Planck-Institut f\"{u}r Astronomie (MPIA Heidelberg), Max-Planck-Institut für Astrophysik (MPA Garching), Max-Planck-Institut f\"{u}r Extraterrestrische Physik (MPE), National Astronomical Observatories of China, New Mexico State University, New York University, University of Notre Dame, Observat\'{o}rio Nacional / MCTI, The Ohio State University, Pennsylvania State University, Shanghai Astronomical Observatory, United Kingdom Participation Group, Universidad Nacional Aut\'{o}noma de México, University of Arizona, University of Colorado Boulder, University of Oxford, University of Portsmouth, University of Utah, University of Virginia, University of Washington, University of Wisconsin, Vanderbilt University, and Yale University.

Some of the observations reported in this paper were obtained with the Southern African Large Telescope (SALT) under the program 2020-2-DDT-004 (PI: E. Niemczura).

Some of the observations reported in this paper were made with ESO Telescopes at the La Silla Paranal Observatory under program ID 105.20DR (PI: T. Rivinius).

This work has made use of data from the European Space Agency (ESA) mission {\it Gaia} (\url{https://www.cosmos.esa.int/gaia}), processed by the {\it Gaia} Data Processing and Analysis Consortium (DPAC, \url{https://www.cosmos.esa.int/web/gaia/dpac/consortium}). Funding for the DPAC has been provided by national institutions, in particular the institutions participating in the {\it Gaia} Multilateral Agreement.

AS gratefully acknowledges support by the Fondecyt Regular (project code 1220610), and ANID BASAL projects ACE210002 and FB210003.

\section*{Data Availability}

The \href{https://archive.stsci.edu/missions-and-data/tess}{TESS}, \href{https://asas-sn.osu.edu/}{ASAS-SN}, \href{https://dr17.sdss.org/infrared/spectrum/search}{SDSS/APOGEE}, and \href{http://archive.eso.org/scienceportal/home}{ESO/VLT} data used in this paper are publically available. The APO/ARCES and SALT/HRS spectra are available upon request from S. Drew Chojnowski.



\bibliographystyle{mnras}
\bibliography{tr16-26} 




\bsp	
\label{lastpage}
\end{document}